\documentclass[twocolumn,showpacs,preprintnumbers,amsmath,amssymb]{revtex4}

\usepackage{graphicx}
\usepackage{subfigure}
\graphicspath{{figs/}{figs_small/}}
\usepackage{bm}

\newcommand{\twod}{\perp} 

\begin{document}


\title{Dynamical Effects and Phase Separation in Thin Films}

\author{Lennon \'O N\'araigh}
\author{Jean-Luc Thiffeault}
\email{jeanluc@imperial.ac.uk}
\affiliation{Department of Mathematics, Imperial College London, SW7 2AZ,
United Kingdom}

\date{\today}

\begin{abstract}
We study phase separation in thin films using the Navier--Stokes Cahn--Hilliard
equations in the lubrication approximation, modeling substrate-film interactions
with a van der Waals potential.  We investigate the thin-film equations numerically
and compare them with experimental results.  We find that the model captures
the qualitative features of real phase-separating fluids, in particular the
tendency of concentration gradients to produce film thinning and surface
roughening.  The ultimate outcome of the phase separation depends strongly
on the dynamical backreaction of concentration gradients on the flow, as
we demonstrate when a shear stress is applied at the film's surface.  When
the backreaction is small, the phase domain boundaries align with the direction
of the imposed stress, while as the backreaction is made larger, the domains
begin to align in the perpendicular direction.
\end{abstract}

\pacs{47.15.gm, 47.55.-t, 64.75.+g}
\maketitle

When a binary fluid is cooled below the critical temperature, the homogeneous
state is energetically unfavourable and the system spontaneously phase-separates
and forms domains rich in either fluid component~\cite{CH_papers,Bray_advphys}.
 Due to the relevance of phase-separating thin films in industrial
applications~\cite{Applications}, many experiments
and numerical simulations focus on understanding how phase
separation is altered if the binary fluid is confined in a thin layer.  We
propose a lubrication approximation based on the coupled Navier--Stokes Cahn--Hilliard
equations to explain the main features of these studies.

Several recent experiments have clarified the different regimes of domain
growth in a binary thin film.  Wang and Composto~\cite{WangH2000}
have identified early, intermediate, and late stages of evolution.    The
early stage comprises three-dimensional domain growth, while the intermediate
stage is characterized by the formation of wetting layers at the film boundaries,
the thinning of the middle layer, and significant surface roughening.  Due
to the thinning of the  middle layer, the sandwich-like structure breaks
up and matter from the wetting layer flows back into the bulk.  Thus, a late
stage is reached, consisting of bubbles coated by thin wetting layers.
 This characterization of the evolution has been seen in other experiments~\cite{ChungH2004,CH_experiments},
 although clearly a variety of behaviors is possible, depending on the wetting
 properties of tnhe mixture.  Our model captures the essential features of
 this evolution, in particular the tendency for concentration gradients to
 promote film rupture and surface roughening.

In a series of papers, Das \emph{et al.}~\cite{Puri2005,Others_by_Puri}
investigate the behaviour of  binary fluids with wetting.  In~\cite{Puri2005}
they specialize to ultra-thin films.  In bulk mixtures, where one of the
fluid components is preferentially attracted to the boundary, a layer rich
in that component may be established there, followed by depletion layer,
and so on.  This so-called spinodal wave propagates into the bulk~\cite{Others_by_Puri}.
 In ultra-thin films, the film thickness is less than a single spinodal wavelength
 and the spinodal wave is suppressed.  Two distinct outcomes of phase
 separation are identified, depending on whether one binary fluid
 component wets the film boundary completely or partially.  Our focus will
 be on the partially wet case.  In this wetting regime, both fluid components
 are in contact with the film boundaries.  The authors
 find an ultimate state of domain formation extending in the lateral directions
 and growing in time as $t^{1/3}$, a result that indicates domain growth
 by Lifshitz--Slyozov diffusion~\cite{LS}.

 These papers elucidate the role of wetting and film thickness on the process
 of phase separation, although they do not discuss hydrodynamics or the effect
 of free-surface variations on domain formation.  In this paper, we therefore
 focus on ultra-thin films with a variable free surface, and for simplicity
 we restrict our attention to the case where both fluids experience the same
 interaction with the substrate and free surface.
%
%
The model we introduce is based on the Navier--Stokes Cahn--Hilliard (NSCH)
equations~\cite{LowenTrus} and gives a qualitative explanation of these studies,
in particular the tendency of domain formation to cause film rupture and
surface roughening.  With an applied external forcing, the model illustrates
the salient effect of the dynamical backreaction of concentration gradients
on the flow, a useful result in applications where control of phase separation
is required~\cite{Krausch1994}.

In full generality, the equations we study are
\begin{subequations}
\begin{gather}
\frac{\partial \bm{v}}{\partial t}+\bm{v}\cdot\nabla\bm{v}=\nabla\cdot\bm{T}-\frac{1}{\rho}\nabla
\phi,\\
\frac{\partial c}{\partial t}+\bm{v}\cdot\nabla c=D\nabla^2\left(c^3-c-\gamma\nabla^2c\right),\\
\nabla\cdot\bm{v}=0,
\end{gather}%
where
\begin{gather}
T_{ij} =-\frac{p}{\rho}\delta_{ij}+\nu\left(\frac{\partial v_i}{\partial
x_j}+\frac{\partial v_j}{\partial x_i}\right)-\beta\gamma\frac{\partial c}{\partial
x_i}\frac{\partial c}{\partial x_j}
\end{gather}
\label{eq:NSCH}%
\end{subequations}%
is the stress tensor, $p$ is the fluid pressure, $\phi$ is the body force potential
and $\rho$ is the constant density.  Additionally, $\nu$ is the kinematic
viscosity, $\beta$ is the mixture free energy per unit mass, $D$ is the
Cahn--Hilliard diffusion coefficient, and $\sqrt{\gamma}$ is the thickness
of domain boundaries.  The concentration boundary condition for Eq.~\eqref{eq:NSCH}
is $\bm{n}\cdot\nabla c = \bm{n}\cdot\nabla\left(c^3-c-\gamma\nabla^2 c\right)=0$,
where $\bm{n}$ is a vector normal to the boundary, while the velocity boundary
conditions on the velocity and stress tensor are standard~\cite{Oron1997}.
 We nondimensionalize these equations by using the vertical length scale
 $h_0$, the horizontal or lateral length scale $\lambda$, and the diffusion
 time $\lambda^2/D$.  If the parameter $\varepsilon=h_0/\lambda$ is small,
 a lubrication approximation is possible~\cite{Oron1997}.  We take the following
 dimensionless groups to be of order unity,
%
%
%
%
\begin{gather*}
Re=\frac{\varepsilon D}{\nu},\qquad  C_{\phantom{}}=\frac{D}{\varepsilon^2h_0^2\sigma_0\rho\nu},\\
r=\frac{\varepsilon^2\beta\gamma}{D\nu},\qquad
C_{\mathrm{n}}=\frac{\varepsilon\sqrt{\gamma}}{h_0},
\end{gather*}
where $Re$ is the Reynolds number, $C_{\mathrm{n}}$ is the Cahn number~\cite{LowenTrus}
which provides a dimensionless measure of domain wall thickness, $r$
is a dimensionless measure of the backreaction strength, and $C^{-1}$ is
a
dimensionless measure of surface tension corresponding to the dimensional
surface tension $\sigma_0$.  Using these scalings, we expand the nondimensional
version of Eq.~\eqref{eq:NSCH} in powers of $\varepsilon$, following the
method outlined in~\cite{Oron1997}, and obtain equations for the free surface
height $h\left(x,y,t\right)$ and concentration $c\left(x,y,t\right)$,
\begin{subequations}
\begin{gather}
\frac{\partial h}{\partial t}+\nabla_{\twod}\cdot\left(\bm{u} h\right)=0,\\
\frac{\partial}{\partial t}\left(c h\right)+\nabla_{\twod}\cdot\left(\bm{u}c
h\right)=\nabla_{\twod}\cdot\left(h\nabla_{\twod}\mu\right),
\end{gather}%
\label{eq:model}%
\end{subequations}%
where
\begin{gather*}
\bm{u}=\tfrac{1}{2}h\nabla_{\twod}\sigma-\tfrac{1}{3}h^2\nabla_{\twod}p,\\
p = -\frac{1}{C}\nabla_{\twod}^2 h+\phi\left(x,y,h\left(x,y,t\right)\right)+r\left(\nabla_{\twod}c\right)^2,\\
\mu=c^3-c-C_{\mathrm{n}}^2\frac{1}{h}\nabla_{\twod}\cdot\left(h\nabla_{\twod}c\right).
\end{gather*}%
Here $\nabla_{\twod}=\left(\partial_x,\partial_y\right)$ is the gradient
operator in the lateral directions, $\sigma$ is the surface tension,
$\phi$ is the body force potential, $\bm{u}$ is a vertically-averaged
velocity, $p$ as a vertically-averaged pressure, and $\mu$ as the chemical
potential.  While the equations do not allow for vertical variations in concentration,
we show in what follows that the model reproduces the qualitative features
observed in thin binary fluids, especially in the case where both
binary fluid components interact identically with the substrate and free
surface~\cite{Puri2005}.

For thin films with $h_0=100$--$1000\text{ nm}$~\cite{ChungH2004,WangH2000},
the dominant contribution to the potential is due to van der Waals interactions~\cite{Oron1997,Book_Parsegian2006},
and following these authors we take $\phi=A/h^3$,
where $A$ is the dimensionless Hamaker coefficient.  To prevent
rupture~\cite{Oron1997}, we study films where $A<0$, and take $A$ to be independent
of the concentration level, so that both binary fluid components are attracted
equally to the substrate and free surface boundaries.  In this case, Eq.~\eqref{eq:model}
possesses simple one-dimensional equilibrium solutions, obtained by setting
$\bm{u}=\nabla_{\twod}\mu=0$.
\begin{figure}[htb]
\subfigure[]{
  \scalebox{0.29}[0.29]{\includegraphics*[viewport=0 0 390 300]{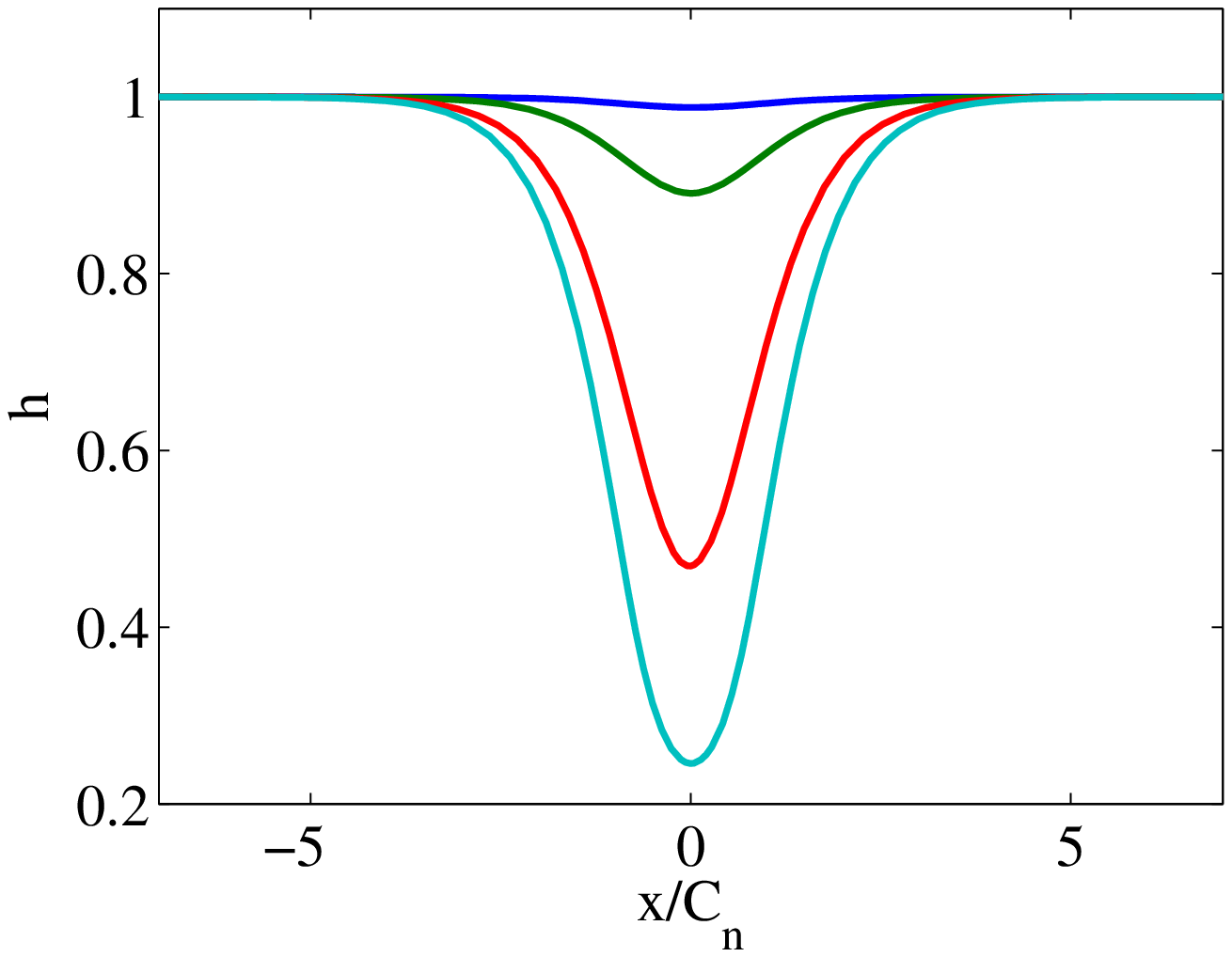}}
}
\subfigure[]{
  \scalebox{0.29}[0.29]{\includegraphics*[viewport=0 0 390 300]{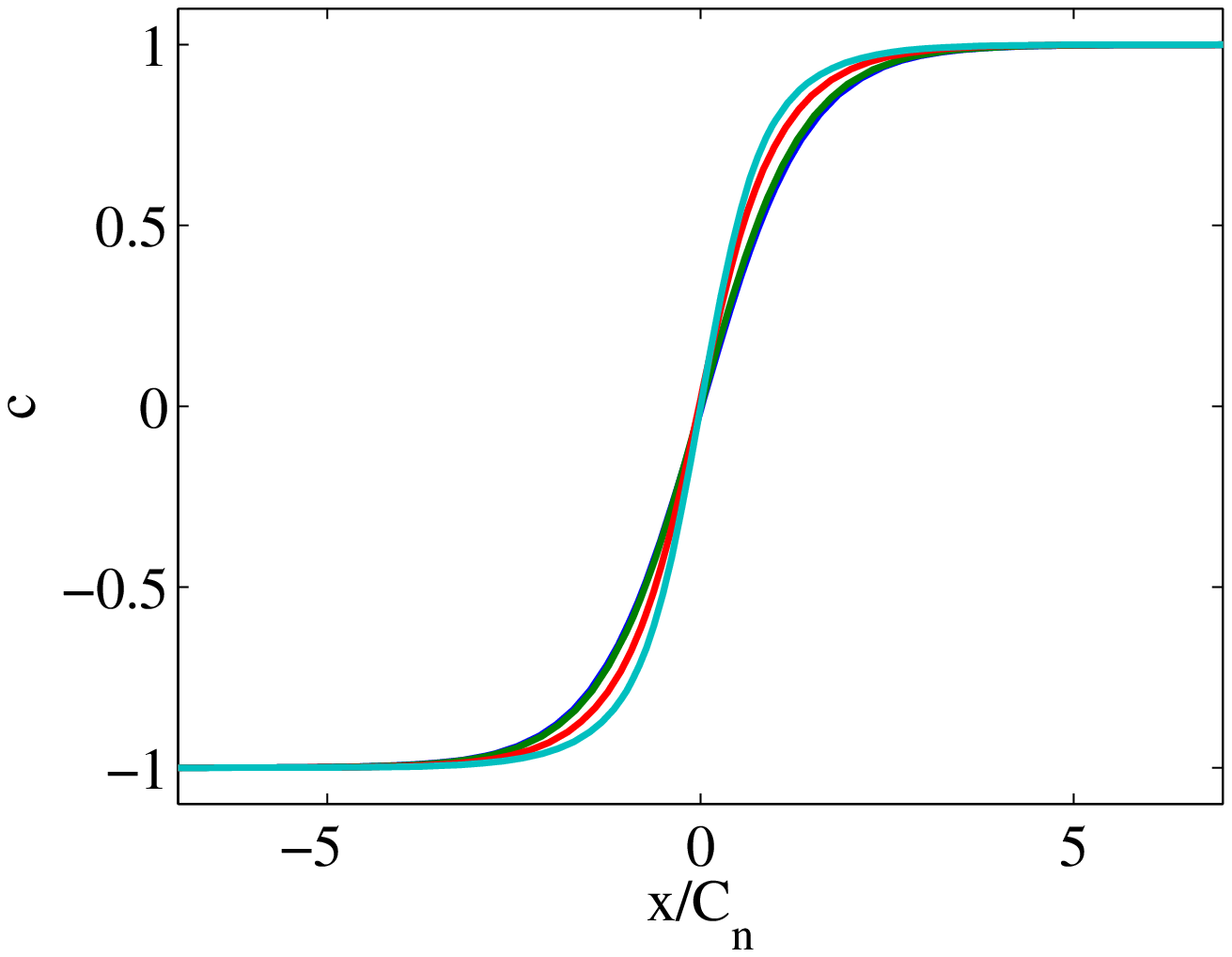}}
}
\caption{(Color online) Equilibrium solutions of Eq.~\eqref{eq:model} for
$C=C_{\mathrm{n}}^2=|A|=1$ and $r=0.1,1,10,50$.
 In (a) the valley deepens with increasing $r$ although the film
 never ruptures, while in (b) the front steepens with increasing $r$.}
\label{fig:eqm}
\end{figure}
From Fig.~\ref{fig:eqm} we see that the one-dimensional equilibrium solution
of Eq.~\eqref{eq:model}, with boundary conditions $h\left(\pm\infty\right)=1$,
$c\left(\pm\infty\right)=\pm1$, consists of a step-like profile for the concentration,
corresponding to a pair of domains separated by a smooth transition region.
 Across this transition region, the height field dips into a valley.  While
 the valley increases in depth for large backreaction strength $r$, the film
 never ruptures.  This result follows from the inequality $h''\left(0\right)>0$,
 since $x=0$ is a local minimum.  Thus, from the equilibrium condition $u=0$,
\[
0<\left[1+\frac{r}{\left|A\right|}c'\left(0\right)^2\right]^{-1}<\left[h\left(0\right)\right]^{3}.
\]
In this way, the repulsive van der Waals potential has a regularizing effect
on the solutions.

Physically, the formation of the valley arises from the balance
between the van der Waals and backreaction effects.  From the solution in
Fig.~\ref{fig:eqm}, the capillary force $F_{\mathrm{cap}}=-r\partial_x\left(\partial_x
c\right)^2$ and the van der Waals force $F_{\mathrm{vdW}}=\left|A\right|\partial_x
h^{-3}$ always have opposite sign.  The repulsive van der Waals force acts
as a non-linear diffusion~\cite{Laugesen2002} and inhibits rupture, and therefore
$F_{\mathrm{cap}}$ promotes rupture, a result seen in experiments~\cite{WangH2000}.
 The valley in the height field represents a balance between the
 smoothening and the rupture-inducing effects.

As in ordinary Cahn--Hilliard dynamics~\cite{Bray_advphys}, the one-dimensional
equilibrium solution hints at the late-time configuration in higher dimensions.
 Thus, we expect the multidimensional solution to comprise concentration
 domains with a height field of peaks
and valleys, with valleys occurring at domain boundaries.  We have verified
with numerical simulations that this is indeed the case.  By using a measure
of domain size $\left(L_x,L_y\right)$ based on the Fourier transform of the
correlation function $\langle c\left(\bm{x},t\right)c\left(\bm{x}+\bm{r},t\right)\rangle$~\cite{ONaraigh2007},
we have found that the domains grow in time as $t^{1/3}$, the usual Lifshitz-Slyozov
growth law~\cite{LS}.  Here $\bm{x}=\left(x,y\right)$ denotes the lateral
coordinates and $\langle...\rangle$ denotes the spatial average.  The modified
growth exponent due to hydrodynamic effects~\cite{Bray_advphys,
Berti2005} is not observed, a result that emerges from the non-linear diffusive
character of the height equation, which damps any undulations not caused
by concentration gradients.  The surface roughness arising from the concentration
gradients is similar to that observed in the one-dimensional case and has
been seen in several experiments~\cite{WangH2000, Jandt1996}.

\begin{figure}[htb]
\subfigure[]{
  \scalebox{0.12}[0.12]{\includegraphics*[viewport=70 0 510 420]{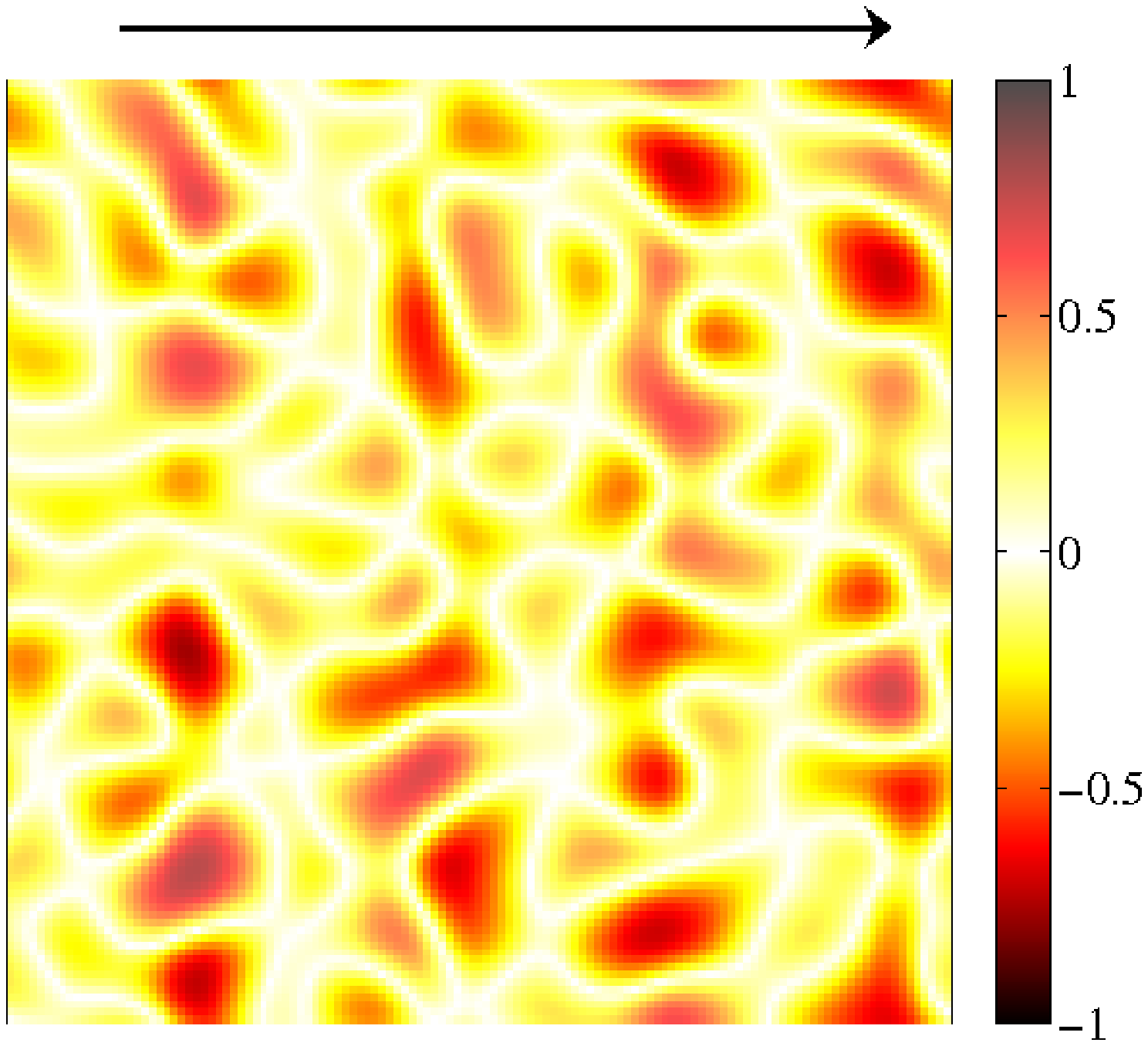}}
}
\subfigure[]{
  \scalebox{0.12}[0.12]{\includegraphics*[viewport=0 0 400 420]{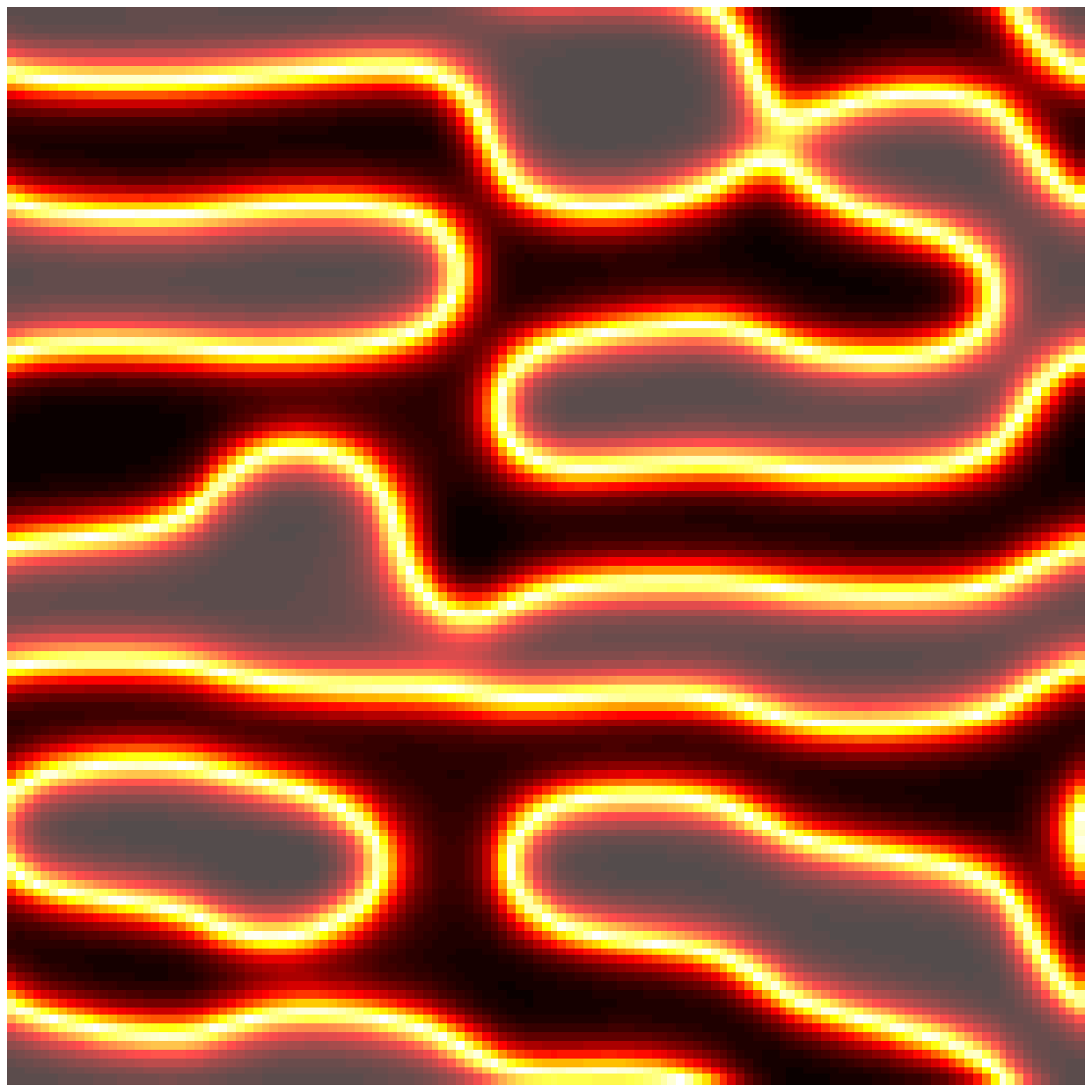}}
}
\subfigure[]{
  \scalebox{0.12}[0.12]{\includegraphics*[viewport=0 0 400 420]{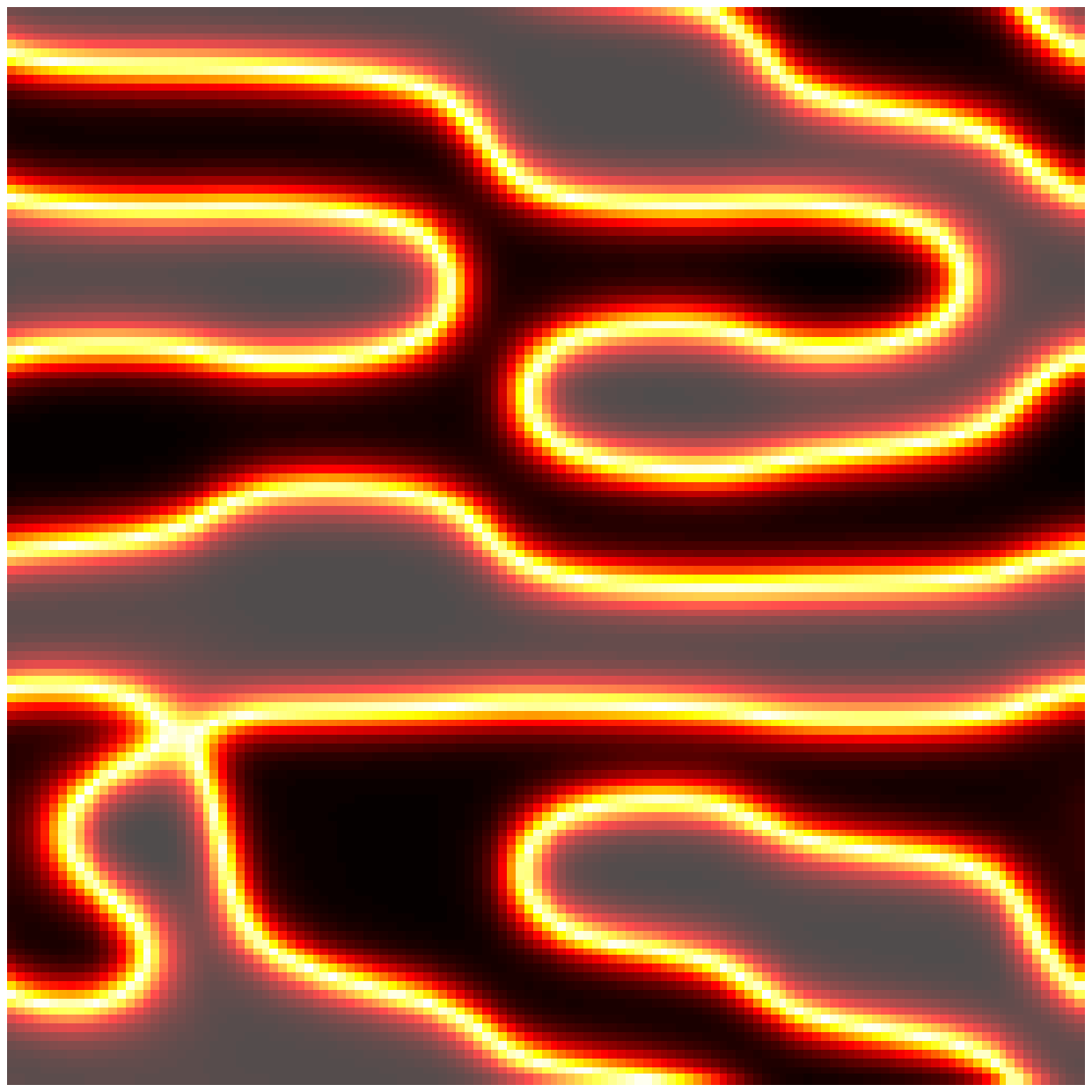}}
}
\subfigure[]{
  \scalebox{0.12}[0.12]{\includegraphics*[viewport=0 0 400 420]{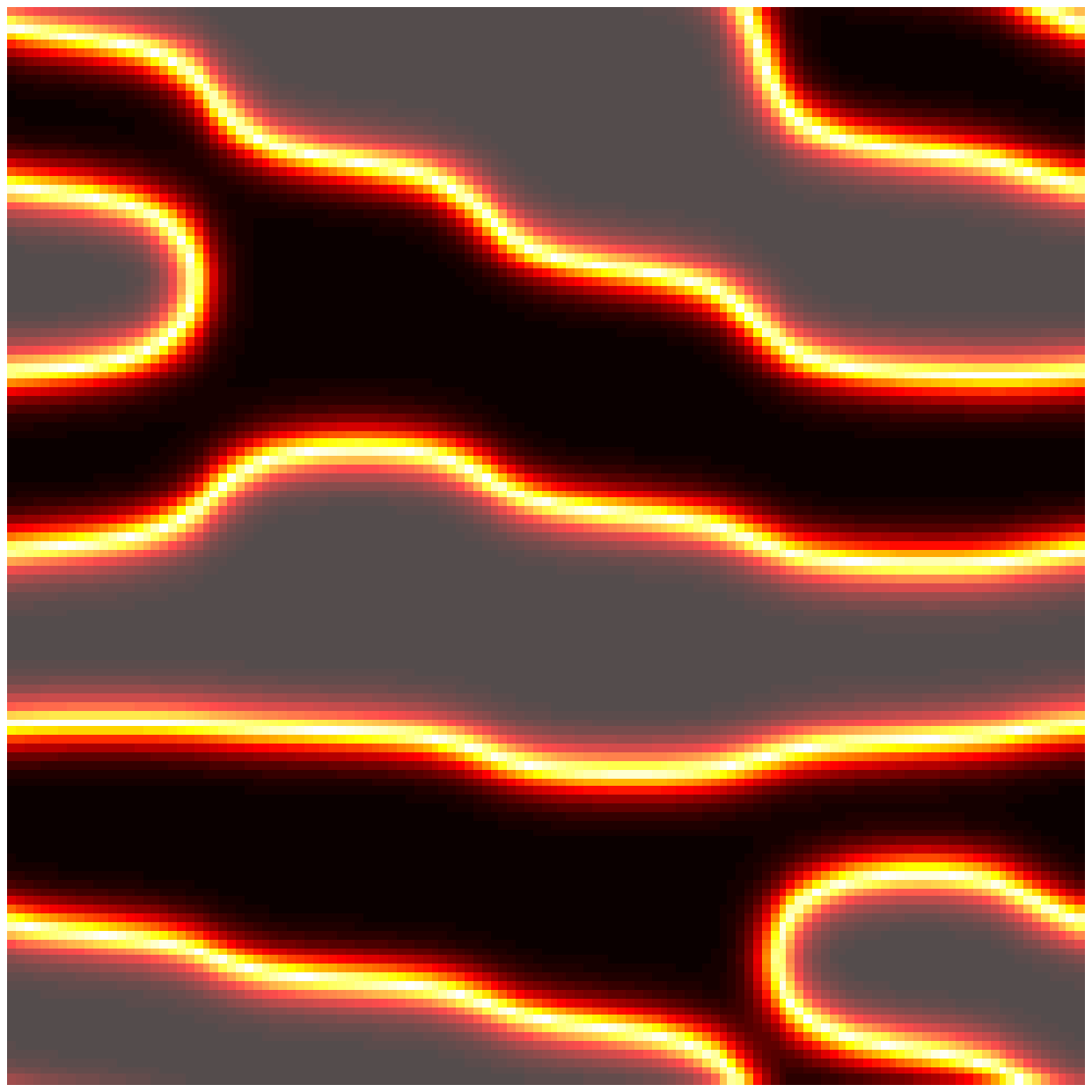}}
}
\subfigure[]{
  \scalebox{0.12}[0.12]{\includegraphics*[viewport=0 0 400 420]{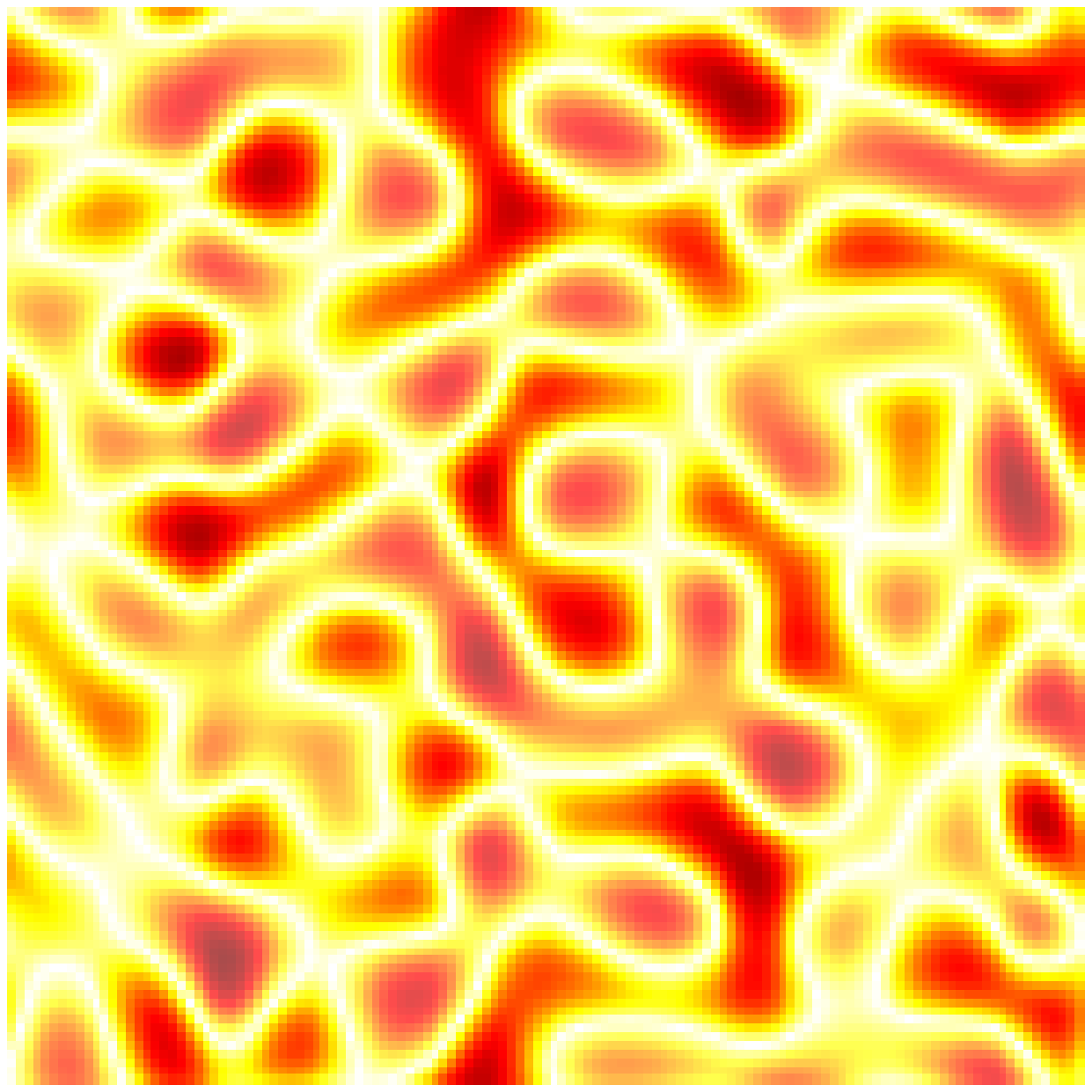}}
}
\subfigure[]{
  \scalebox{0.12}[0.12]{\includegraphics*[viewport=0 0 400 420]{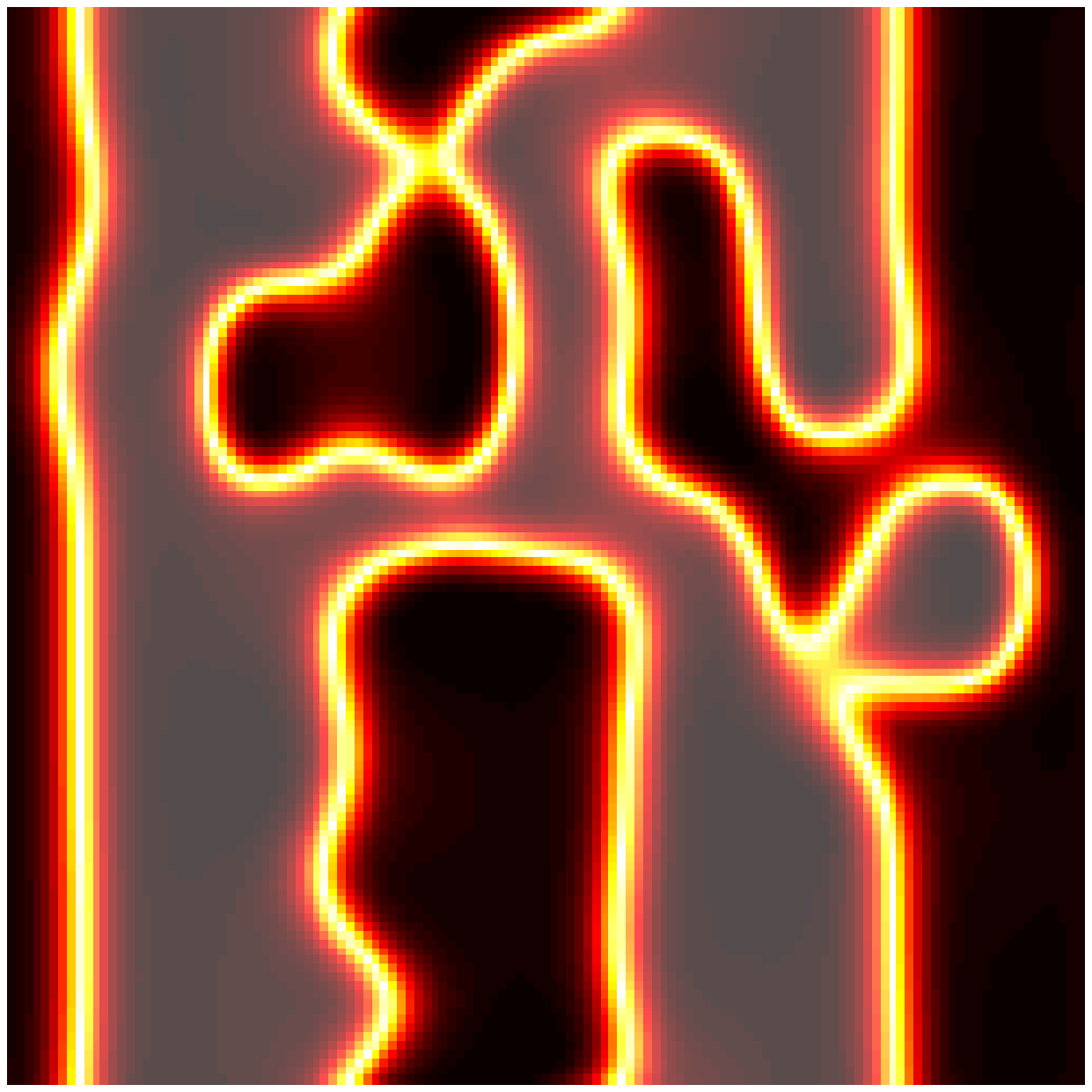}}
}
\subfigure[]{
  \scalebox{0.12}[0.12]{\includegraphics*[viewport=0 0 400 420]{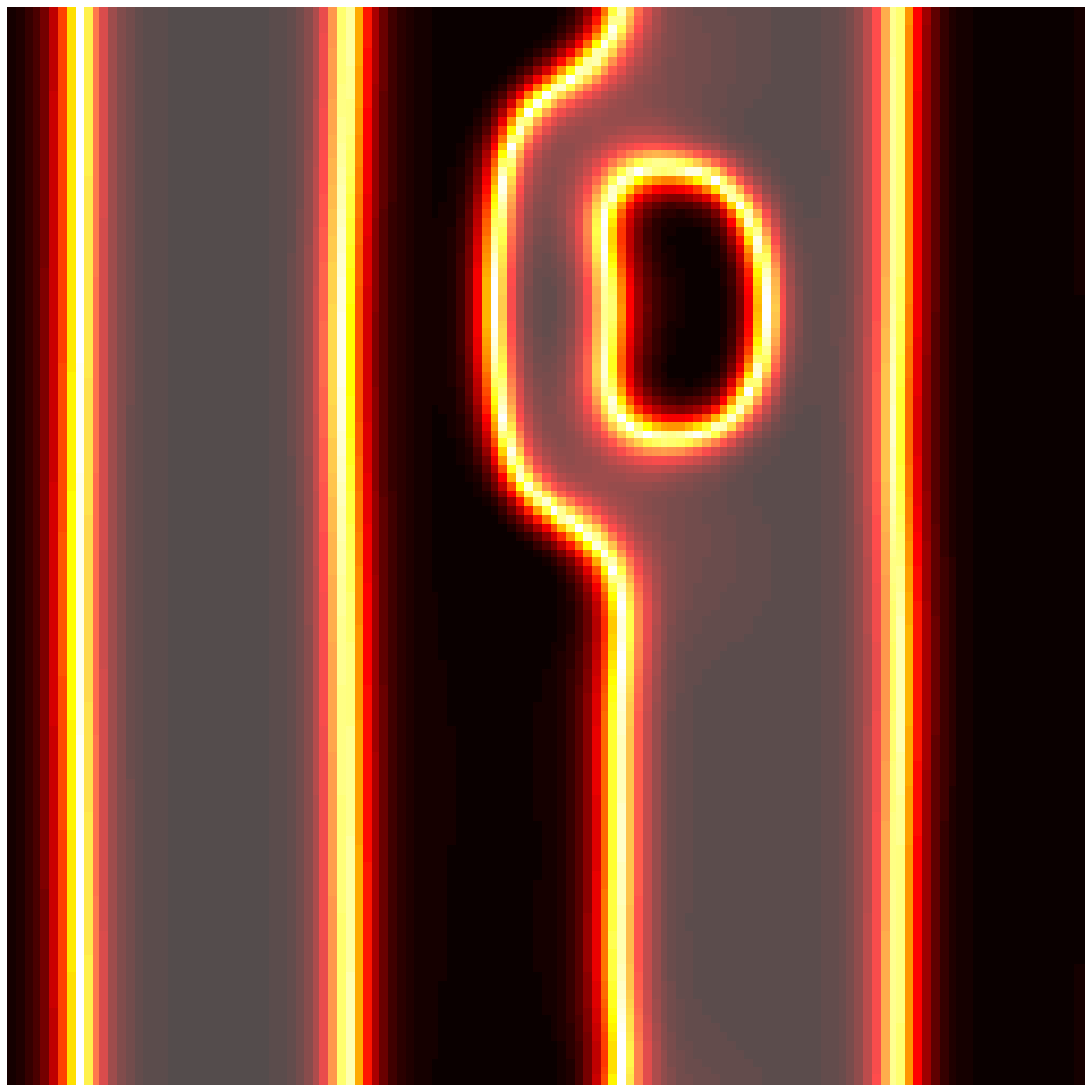}}
}
\subfigure[]{
  \scalebox{0.12}[0.12]{\includegraphics*[viewport=0 0 400 420]{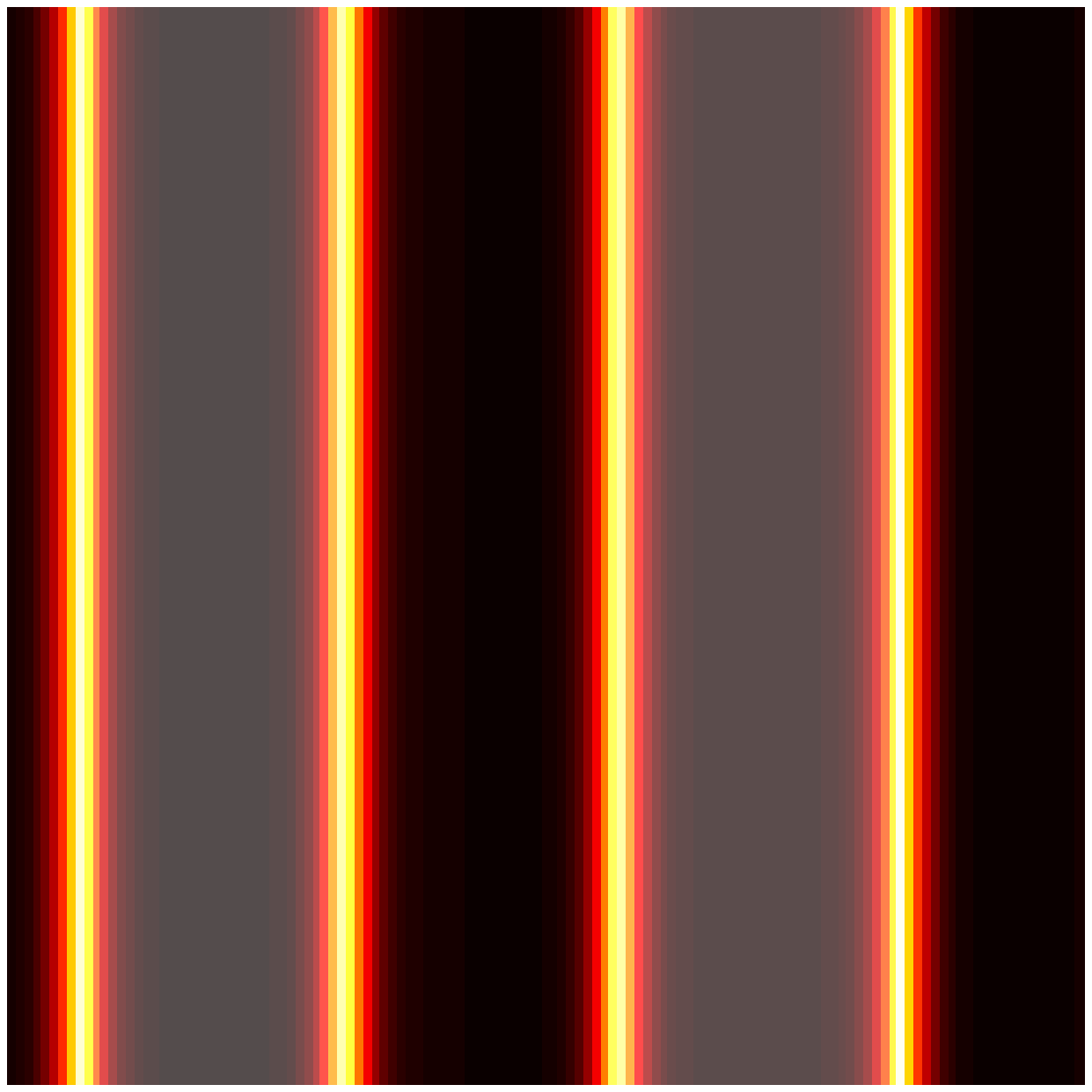}}
}
\caption{(Color online) The concentration field for $C=-A=1$.  Across the
first row, $r=0$ and (a) $t=500$; (b) $t=3750$; (c) $t=7500$; (d) $t=30000$.
 Across the second row, $r=\tfrac{1}{2}$ and (e) $t=500$; (f) $t=3750$; (g)
 $t=7500$; (h) $t=30000$.  The surface tension gradient is parallel to the
 arrow and $\sigma=\sigma_0\sin\left(k x\right)$, $\sigma_0=20$
 and $k=4k_0$. In Figs.~(a)--(d) with $r=0$, the domains align along the
 arrow, while in Figs.~(e)--(h) with moderate backreaction strength, the
 domains align in a direction perpendicular to the arrow.}
\label{fig:domains}
\end{figure}

The dramatic effect of the reaction of the concentration gradients on the
phase separation is apparent when we apply a surface tension gradient across
the film.  Physically, this can be realized by differential heating of
the surface~\cite{Heating_surface}, although a surfactant will also induce
stresses at the surface~\cite{Bush2001}.  We set $\sigma=\sigma_0\sin
k x$, where $k=\left(2\pi/L\right)m = k_0 m$ is the spatial scale of the
 surface tension variation and $m$ is an integer. Then the velocity that
 drives the system becomes
\begin{multline}
\bm{u}=\tfrac{1}{2}h\left(k\sigma_0\cos k x,0\right)\\+\tfrac{1}{3}h^2\nabla_{\twod}\left[\frac{1}{C}\nabla_{\twod}^2
h+\frac{|A|}{h^3}-r\left(\nabla_{\twod} c\right)^2\right].
\label{eq:velocity_st}
\end{multline}
This velocity field may also be obtained by imposing a shear stress $\bm{\tau}$
at the surface, provided $\bm{\tau}=\nabla\sigma$~\cite{Myers2002}.  We carry out simulations
using Eq.~\eqref{eq:velocity_st} on a $128\times128$ grid.  The results do
not change upon increasing the resolution.  We choose $C_{\mathrm{n}}$
so that domain boundaries are resolved.  The other parameter
values are indicated in the caption to Fig.~\ref{fig:domains}.

\begin{figure}[htb]
  \scalebox{0.45}[0.45]{\includegraphics*[viewport=30 80 400 250]{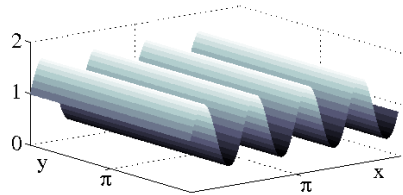}}
\caption{The height field for $r=0$ and $t=30000$ aligns with the applied
surface tension.  The height field at $t=30000$ for $r=\tfrac{1}{2}$ is similar.}
\label{fig:height_st}
\end{figure}

\begin{figure}[htb]
\subfigure[]{
 \scalebox{0.28}[0.26]{\includegraphics*[viewport=0 0 400 340]{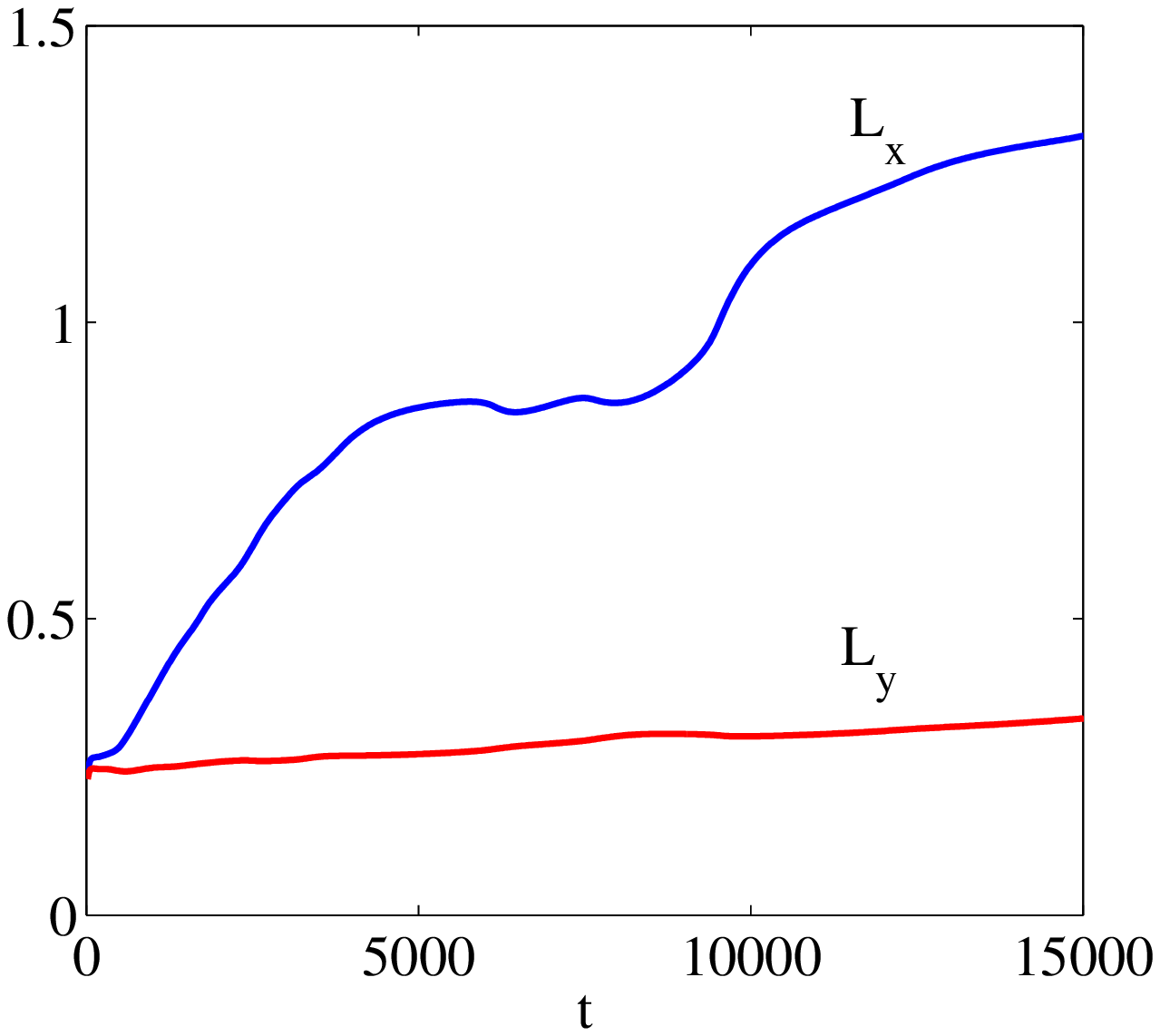}}
}
\subfigure[]{
 \scalebox{0.28}[0.26]{\includegraphics*[viewport=0 0 400 340]{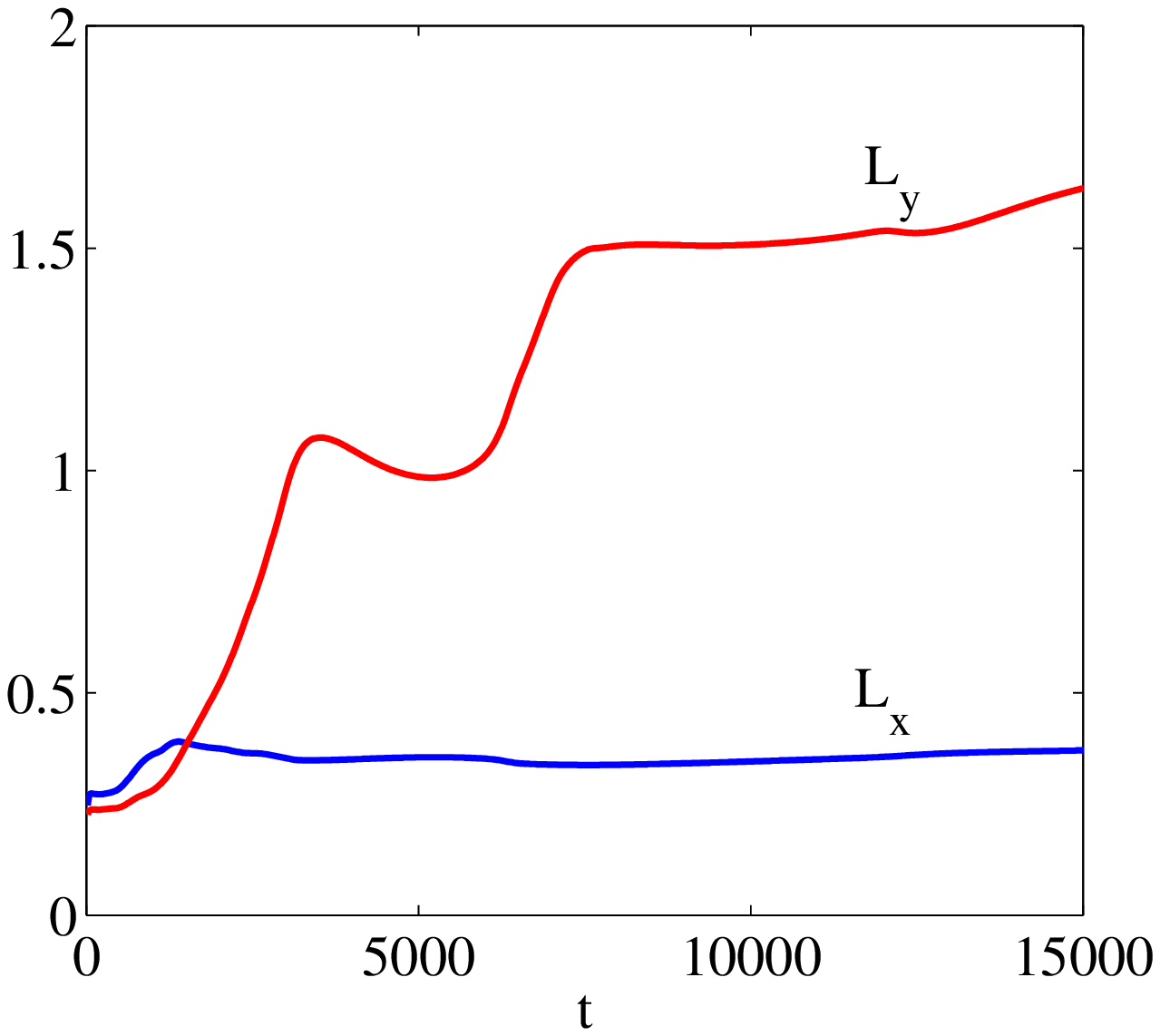}}
}
\caption{(Color online) Growth of $L_x$ and $L_y$ for (a) $r=0$, where $L_x$
grows and $L_y$ saturates or undergoes slow growth.  Since the growth rate
of $L_y$ is small, it is not clear whether saturation or slow linear growth
takes place; (b) $r=\tfrac{1}{2}$, where $L_x$ saturates and $L_y$ grows.}
\label{fig:scales_st}
\end{figure}

This choice of velocity field leads to control of phase separation in the
following manner.  For small values of the backreaction strength, with $r\rightarrow0$,
the height field quickly aligns with the surface tension profile as in Fig.~\ref{fig:height_st},
since the strong effect of the van der Waals diffusion destroys the unforced
part of $h\left(\bm{x},t\right)$. At the same time, the concentration field
begins to form domains.  At later times, when $L_x(t),L_y\left(t\right)\sim2\pi/k$,
the domains align with the gradient of the forcing term.   The growth of
the domains continues in this direction and is arrested (or slowed down considerably)
in the direction perpendicular to the forcing.  The domains are string-like,
with kinks occurring along lines where $\sigma\left(x,y\right)$
is minimized, as evidenced by Fig.~\ref{fig:domains}~(a)--(d).  The growth
of $L_x$ and $L_y$ is shown in Fig.~\ref{fig:scales_st}.  It is not clear
whether $L_y$ is arrested or undergoes slow linear growth and so we do not
report its growth rate.

For moderate values of the backreaction strength with $r\sim O(1)$, the height
field again assumes a profile aligned with the surface tension, while
domains of concentration now
align in a direction perpendicular to the forcing gradient.  Domain growth
continues in the perpendicular direction and is arrested in the direction
of the driving-force gradient.  A pattern
of string-like domains emerges, with domain boundaries forming along lines
where both $\sigma\left(x,y\right)$ and $h\left(x,y,t\right)$ are maximized.
Eventually, the domain boundaries align perfectly with the surface tension
maxima, as evidenced in Fig.~\ref{fig:domains}~(e)--(h).  


The control of phase separation by surface shear therefore depends crucially
on the backreaction.
 This result is amplified by the existence of a no-rupture condition only
 for the $r=0$ case (no backreaction).  This condition relies
 on the alignment of the height and surface tension profiles, which is exact
 only when the backreaction is zero.  Then, at late times, the system evolves
 towards equilibrium and is described by the steady state $\nabla_{\twod}\cdot\left(\tfrac{1}{2}h^2\nabla_{\twod}\sigma-\tfrac{1}{3}h^3\nabla_{\twod}p\right)=0$,
 which by the alignment property reduces to the one-dimensional equation
\[
h^2\left[\tfrac{1}{2}\frac{d\sigma}{d x}+\tfrac{1}{3}h\frac{d}{d
x}\left(\frac{1}{C}\frac{d^2h}{d x^2}+\frac{\left|A\right|}{h^3}\right)\right]=\mathrm{const.}
\]
By multiplying both sides of the expression by
$h$, differentiating and then evaluating the result at $x_0$, a minimum of
both surface tension and height, we obtain the condition
\begin{equation}
\bigl[h\left(x_0\right)\bigr]^3\left[\frac{1}{3C}h\left(x_0\right)h^{\left(4\right)}\left(x_0\right)+\tfrac{1}{2}k^2\sigma_0\right]=\left|A\right|h''\left(x_0\right).
\label{eq:no_rupture_st}
\end{equation}
Since $x_0$ is a minimum of height, $h''\left(x_0\right)>0$, which prevents
$h\left(x_0\right)$ from being zero.  On the other hand, for $r$ and $\sigma_0$
sufficiently large, the alignment of height and surface tension profiles
is not exact, the one-dimensional state is never reached and hence the result
in Eq.~\eqref{eq:no_rupture_st} does not apply.  In that case, simulations
show that the film ruptures in finite time.
  
Given an applied surface tension gradient, we have outlined, by numerical
simulations and calculations, three possible outcomes for the phase separation,
depending on the backreaction strength $r$.  For $r\ll1$, the concentration
forms string-like domains, aligned with the applied force.  For $r\sim O\left(1\right)$,
the concentration forms
domains that align perfectly in a direction perpendicular to the applied
force.  For $r\gg1$, the forcing causes the film to rupture.  The interfacial
tension or backreaction must therefore be chosen carefully in a real fluid
to achieve the desired outcome.

In conclusion, we have derived a thin-film model of phase separation based
on the Navier--Stokes Cahn--Hilliard equations, in which the reaction of
concentration gradients on the flow is important.  We have used this model
to give a qualitative picture of the features of phase separation in real
thin films, in particular the tendency of concentration gradients to promote
rupture in the film, and to produce peaks and valleys in the free surface
that mirror the underlying domain morphology.  We have found that in the
presence of a unidirectional sinusoidal variation in surface tension, the
strength of the backreaction determines the direction in which the domains
align.  This result could prove useful in microfabrication applications where
control of phase separation is required~\cite{Krausch1994}.

Because the lubrication model suppresses vertical variations in the concentration
field, we are limited to the case where the binary fluid components interact
identically with the boundaries at the substrate and free surface.
 However, the model quite generally gives an accurate description of surface
 roughening arising from van der Waals forces.  More detailed models based
 on this approach, involving different boundary conditions that better reflect
 wetting behaviour~\cite{Puri2005,Racke2003} and a concentration-dependent
 Hamakar coefficient, will capture a wider range of thin-film behaviour.

L.O.N. was supported by the Irish government and the UK Engineering and Physical
Sciences Research Council. J.-L.T. was supported in part by the UK EPSRC
Grant No. GR/S72931/01.



\begin{thebibliography}{10}

\bibitem{CH_papers}
J.~W. Cahn and J.~E. Hilliard, 
\newblock {\em J. Chem. Phys}, 28:258--267, 1957;
%
%
J.~Zhu, L.~Q. Shen, J.~Shen, V.~Tikare, and A.~Onuki, \newblock {\em
Phys. Rev. E}, 60:3564--3572, 1999.

\bibitem{Bray_advphys} 
A.~J. Bray, 
\newblock {\em Adv. Phys.}, 43:357--459, 1994.

\bibitem{Applications}
A.~Karim, J.~F. Douglas, L.~P. Sung, and B.~D. Ermi, \newblock in {\em Encyclopedia
of Materials: Science and Technology} (Elsevier, Amsterdam, 2002); 
%
%
D.~L. Smith, \newblock {\em Thin-Film Deposition: Principles and Practice},
(McGraw-Hill, New York, 1995); K.~Mertens, V.~Putkaradze, D.~Xia,
and S.~R. Brueck, \newblock {\em J. App. Phys.}, 98:034309, 2005.

\bibitem{WangH2000}
H.~Wang and R.~J. Composto,
\newblock {\em J. Chem. Phys.}, 113:10386, 2000.

\bibitem{ChungH2004}
H.~Chung and R.~J. Composto,
\newblock {\em Phys. Rev. Lett.}, 92:185704--1, 2004.

\bibitem{CH_experiments}
W.~Wang, T.~Shiwaku, and T.~Hashimoto,
\newblock {\em Macromolecules}, 36:8088, 2003;
%
%
J.~Klein H.~Hoppe, M.~Heuberger,
\newblock {\em Phys. Rev. Lett.}, 86:4863, 2001.

\bibitem{Puri2005}
S.~K. Das, S.~Puri, J.~Horbach, and K.~Binder,
\newblock {\em Phys. Rev. E}, 72:061603, 2005.

\bibitem{Others_by_Puri}
S.~Puri and K.~Binder, \newblock {\em Phys. Rev. E}, 66:061602, 2002; 
%
%
S.~Puri and K.~Binder, \newblock {\em Phys. Rev. Lett.}, 86:1797, 2001;
%
%
S.~Puri, K.~Binder, and H.~L. Frisch, \newblock {\em Phys. Rev. E}, 56:6991,
1997.

\bibitem{LS}
I.~M. Lifshitz and V.~V. Slyozov,
\newblock {\em J. Chem. Phys. Solids}, 19:35--50, 1961.

\bibitem{LowenTrus}
J.~Lowengrub and L.~Truskinowsky,
\newblock {\em Proc. R. Soc. London, Ser. A}, 454:2617--2654, 1998.

\bibitem{Krausch1994}
G.~Krausch, E.~J. Kramer, M.~H. Rafailovich, and J.~Sokolov,
\newblock {\em Appl. Phys. Lett.}, 64:2655, 1994.

\bibitem{Oron1997}
A.~Oron, S.~H. Davis, and S.~G. Bankoff,
\newblock {\em Rev. Mod. Phys.}, 69:931, 1997.

\bibitem{Book_Parsegian2006}
V.~A. Parsegian,
\newblock {\em Van der Waals Forces} (Cambridge University Press, New York, 2001).

\bibitem{Laugesen2002}
R.~S. Laugesen and M.~C. Pugh,
\newblock {\em Electron. J. Diff. Eqns.}, 2002:1, 2002.

\bibitem{ONaraigh2007}
L.~\'O N\'araigh and J.-L. Thiffeault,
\newblock {\em Phys. Rev. E}, 75:016216, 2007.

\bibitem{Berti2005}
S.~Berti, G.~Boffetta, M.~Cencini, and A.~Vulpiani,
\newblock {\em Phys. Rev. Lett.}, 95:224501, 2005.

\bibitem{Jandt1996}
K.~D. Jandt, J.~Heier, F.~S. Bates, and E.~J. Kramer,
\newblock {\em Langmuir}, 12:3716, 1996.

\bibitem{Heating_surface}
A.~P. Krekhov and L.~Kramer,
\newblock {\em Phys. Rev. E}, 70:061801, 2004;
%
%
N.~Garnier, R.~O. Grivoriev, and M.~F. Schatz,
\newblock {\em Phys. Rev. Lett.}, 91:054501, 2003.

\bibitem{Bush2001}
A.~E. Hosoi and J.~W.~M. Bush
\newblock{\em J. Fluid Mech.}, 442:217, 2001.

\bibitem{Myers2002}
T.~G. Myers, J.~P.~F. Charpin, and C.~P. Thompson,
\newblock{\em Phys. Fluids}, 14:240, 2001.

\bibitem{Racke2003}
R.~Racke and S.~Zheng,
\newblock {\em Adv. Dil. Equations}, 8:83, 2003.

\end{thebibliography}
\end{document}